# DevOps: A Historical Review and Future Works

Mayank Gokarna[a], Raju Singh[b]

[a] IBM India Pvt Ltd, Subramanya Arcade, Bangalore, India
[b] IBM India Pvt Ltd, Manyata Tech Park, Bangalore, India

A B S T R A C T

DevOps is an emerging practice to be followed in Software Development life cycle. The name DevOps indicates that it's an integration of Development and Operations team. It is followed to integrate the various stages of the development lifecycle. DevOps is an extended version of the existing Agile method. DevOps aims at Continuous Integration, Continuous Delivery, Continuous Improvement, faster Feedback and Security. This paper reviews the building blocks of DevOps, challenges in adopting DevOps, Models to improve DevOps practices and Future works on DevOps.

## 1. Introduction:

The practices followed during the software development lifecycle plays an important role. In the conventional development lifecycle, different teams will play their role at their own level. Separate teams make the product life cycle lengthier and also the communication between the teams poor (Lwakatare et al., 2015). This kind of development model is called the waterfall model. To break the walls between the teams and to enhance the dissemination of the information the new methodology Agile was discovered. Agile means "to move fast and easy". Agile process methodology improved the interaction between individual teams and improved collaboration. Some of the agile principles are Scrum, Extreme Programming, Lean, Kanban and out of these the Scrum was the first developed principle (Saliya Sajith Samarawickrama and Indika Perera 2017). Even though the agile process reduced the time of the development life cycle, there were some gaps that needed to be bridged. This is why DevOps evolved and it is the extension of the agile process. DevOps unifies the Development and Operations team. Automation also plays an important role in DevOps. The processes like maintenance and testing were already automated in Agile (Michael Hüttermann 2012). DevOps cannot be implemented at one stretch, it needs to be implemented step by step or iteratively. DevOps was divided into 4 areas (Michael Hüttermann 2012) and it is shown in figure (Christof Ebert et al., 2016). This paper mainly deals with DevOps and its building blocks, how it's getting improved, adoption methodology, other parameters that help to improve and challenges in implementing the DevOps.

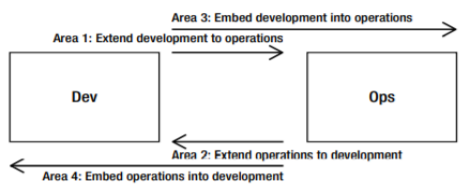

**Fig. 1.** Areas in DevOps

## 2. Agile Methodology:

Due to the shortcomings of the waterfall method, the Agile method was introduced. The agile methodology was introduced in the year 2001 (Strode et al., 2009). The agile is the evolutionary development model (Nerur et al., 2005) and it aims at continuous improvement of the product features. The agile method integrates the programmers, testers, and QA as the Development team and separate Operations team. Conflicts arise between the Development and Operations teams while deploying the newly developed features and fixing problems (Michael Hüttermann 2012). It was hard to maintain the software and to update them whenever necessary (Rodríguez et al., 2013). The agile processes are feature and people-centric approaches and so it was a challenge to change from traditional process-centric approach (Nerur et al., 2005). In the agile method, the information passing from the Development team and operations team was not faster and not in a frequent manner. This is a bottleneck in the agile method (Hemon et al., 2020). This gives rise to the new methodology DevOps, which combines the Development and Operations team to enhance communication and frequency.

## 3. DevOps:

DevOps is a set of procedures which combines the process of Development and Operations (Christof Ebert et al., 2016). DevOps needs a set of tools to perform the function of combination and integration. In other words, DevOps is a single team that looks after development, testing, and operations. In DevOps, the total product cycle doesn't break at any point (Christof Ebert et al., 2016). The DevOps has four Dimensions (Lwakatare et al., 2015), they are

1. Collaboration,
2. Automation,
3. Measurement,
4. Monitoring

DevOps is the extension of the agile method of software development (Jabbari et al., 2016) . DevOps focuses on the continuous delivery of the software along with continuous integration (Jabbari et al., 2016 and Manish Virmani, 2015). Automation also plays a vital role in reducing the latency of product releases. DevOps not only improves collaboration and communication but also fast and continuous delivery, regular updates, increases reliability, etc (Samer Mohamed 2015).

*3.1 Continuous Deployment:*

Cloud computing plays an important role in case of Continuous Delivery (Jabbari et al., 2016). The cloud-based tools help to bridge the gaps between the need and delivery and also provides faster feedback (Manish Virmani, 2015). In case of rapid deployment process, testing should be automated to reduce the latency (Len Bass 2017). In a study (Rahman et al., 2015) done on 19 companies, it was reported that 11 out of 19 companies use Continuous Deployment strategy. The traditional methods like incremental/ iterative and ad-hoc approaches failed to satisfy the needs of the software firms (Sikender Mohsienuddin Mohammad 2019). In (Edwards 2014), the author discussed how to prepare a firm for the continuous delivery process. In that, it was reported that continuous delivery should be adopted as small increments and repetitive tasks should be automated. The benefits of adopting continuous delivery are accelerated time to market, building the right product, improved productivity and efficiency, reliable releases, improved product quality and customer satisfaction (Chen et al., 2017). The code infrastructure helps in fast releases (Michael Hüttermann 2012). Before the emergence of DevOps, there was no shared infrastructure for software development in a firm and so the employees used to work individually on their environment and so the delivery process gets delayed due to increased time for integration.

*3.2 Continuous Integration:*

DevOps itself represents the integration between Development and Operations teams. DevOps enables the continuous integration of all processes involved in product development and so all the process is done by a single team throughout the cycle (Christof Ebert et al., 2016). Inadequate communication is the key problem that triggers the need for DevOps to be implemented. DevOps enables the software firm to provide more features and continuously improve it based on the feedback within a short period (Leah et al., 2016). DevOps not only integrates the teams, but also the tools used in various stages of product development (Bruneo et al., 2014). The new framework was proposed in (Saliya Sajith Samarawickrama and Indika Perera 2017), to implement the process of continuous integration and deployment over the existing Scrum product cycle. The versions of the software tools should be maintained to be compatible with other tools. The version control strategy was proposed to control different artefacts like Source code, configuration files, deployment scripts and binary code of the application in (Nicolás Paez 2018). Automation is necessary for continuous integration and certain tools like Git repository are used to keep track of the changes (Schaefer et al., 2013). HARNESS is a multi-partner research project, which aims at integrating multiple non-homogeneous resources by using the version control techniques (Mark Stillwell and Jose GF Coutinho 2015). A new tool was proposed in (Pérez et al., 2015) called Filling Gap (FG) tool which bridges Development and Operations teams gap, and provides quick feedback on the performance to the Development team.

*3.3 Cloud:*

Cloud is a key player in the DevOps methodology (Borgenholt et al., 2013). The cloud DevOps methodology, tools and culture are shown in figure 2.

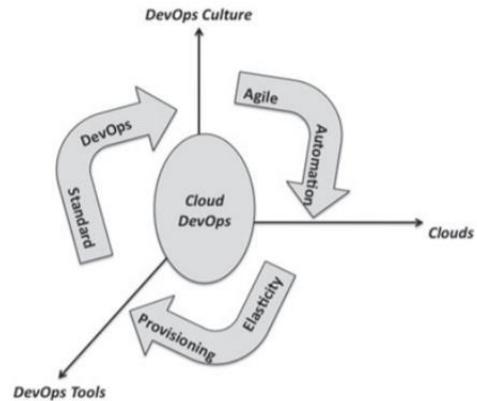

**Fig. 2.** Cloud DevOps (Mohammed Airaj 2017)

The cloud ecosystem enables us to have a large number of interconnected components and easy to access and control the components (Syed and Eduardo 2016) and a simplified cloud ecosystem was proposed. The serverless model uses Cloud in its architecture (Vitalii Ivanov and Kari Smolander 2018). The qualitative data showed that DevOps practices are strongly affected by the new cloud computing model (Vitalii Ivanov and Kari Smolander 2018). For cloud applications, lightweight languages were used (Shigeru Hosono 2012). Cloud helps in iterative software development, monitoring the running applications on the client-side and to get feedback about the process (Bruneo et al., 2014). Use of cloud environment also increases the security of the software applications (Wouter van der Houven 2020). The DevOps artefacts are classified as Node-centric artefacts and Environmental centric artefacts. The TOSCA was used in cloud applications to make the artefacts interoperable (Wettinger et al., 2016).

*3.4 Security:*

DevSecOps or SecDevOps are the terms associated with Secured DevOps. The continuous security model was proposed in (Rakesh Kumar and Rinkaj Goyal 2020), which used open-source software over the cloud to provide security throughout the product development. The security practices used in DevOps (Rahman et al., 2016) are 1)Automation Activities including automated testing, monitoring, code review, software-defined firewall and software licensing 2) Increased collaboration between development and security teams 3) Non-automated security activities are security requirements analysis, performing security configurations, performing security policies, design review, input validation, risk analysis, etc. A Dynamic model was proposed in (Rios et al., 2017) to rectify the security issues in both development and operational activities of the multi-cloud application and the proposed model was validated using the real-time application. The framework of the MUSA Dynamic model is shown in figure 3. The cloud security options include serverless computing, Infrastructure as code and security centralisation (Wouter van der Houven 2020).

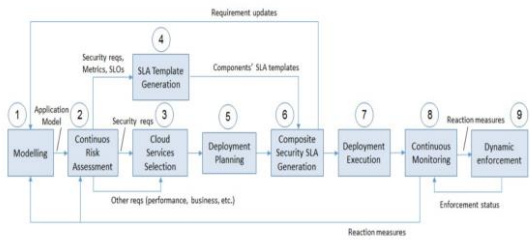

**Fig. 3.** The framework of MUSA DevOps (Rios et al., 2017)

It was stated in (Wouter van der Houven 2020) that cloud makes release faster and more secure. Some of the security activities to be followed in DevOps are Security Requirements Gathering, Threat Modeling, Environment Configuration, Secure Static Analysis, Security-Focused Code Review, Software Penetration Testing, Environment Testing and security Review (Yasar and Kontostathis 2016).

*3.5 Automation in DevOps:*

In case of rapid deployment, the manual testing is impossible and so the testing should be automated. The final task in the automation is to check for errors in the system and to rectify them (Len Bass 2017). Apart from testing, automation also helps in scaling of the products (Schaefer et al., 2013), the network was formed based on the common services that each system would require and specific tools were used for every task. In the proposed model (Arulkumar and Lathamanju 2019), the workflow is divided into backend and frontend workflow. The proposed model tries to automate the processes that are involved in the development lifecycle. In this model, the cycle of the process starts with gathering data and the gathered data is used to build the application and output of the build are tested and then deployed through the cloud. The automation of the software release causes reduced operational cost, increased productivity, increased accessibility & reliability and optimized performance (Sikender Mohsienuddin Mohammad 2019). The Robotic Process Automation (RPA), is an emerging method to automate the repetitive task (Sikender Mohsienuddin Mohammad 2019). Machine learning can be used in automation tasks like testing, quality assurance, fault detection, etc (Kiyana Bahadori and Tullio Vardanega 2018). Automation is the only way to make the release cycle shorter. The architecture was built using reliable existing tools in (Borgenholt et al., 2013) for automated testing and quality assurance. The building blocks of the proposed architecture are Cloud platform, Virtual management machine and configuration management tool. Not every process can be automated, some processes should be done manually. Automation enables fast feedback from the customer side as well (Michael Hüttermann 2012).

*3.6 Tools used in DevOps:*

The tools are required to develop and integrate the processes and it is classified as Build tools and Continuous integration tools (Christof Ebert et al., 2016). Some of the tools for improving DevOps practices are JIRA, GIT, Jenkins and Docker. The tools should be used based on the hierarchy (Laukkarinen et al., 2018). The Chef Cookbooks, Puppet Modules, Saltstack modules, Docker images, Juju charms, Bundles and templates are some of the configuration management tools (Wettinger et al., 2016). Earlier mercurial repository was used to store configuration data and to keep track of the changes (Schaefer et al., 2013). The various tools used in various stages of the DevOps cycles are shown in figure 4. Docker is container based technology and is used to provide isolation between various applications. Using Docker, each application can be configured with specific version of the Operating system running on the host machine. The end application is delivered as the Docker container (Morris et al., 2017). Docker was built on top of the LXC technology, which relies on Linux and hence Docker relies on certain Linux features (Robert Sandoval 2016).

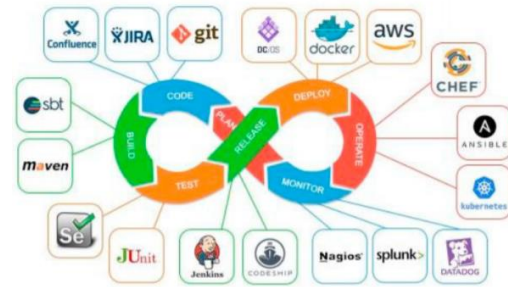

**Fig. 4.** Tools used in DevOps (Arulkumar and Lathamanju 2019)

The project management tool is necessary for an IT firm to manage data of various projects. One such tool is GZ-Agile Project Management Consolidator, which stores and tracks the activities of DevOps as well as Agile (Doukoure et al., 2018). In a fraud detection system developed by Netfective Technology, SimTool was used to evaluate the metrics (Perez-Palacin et al., 2017). Some of the cloud computing services used in DevOps are Amazon Web Services (AWS), Microsoft Azure, IBM Cloud (Arulkumar and Lathamanju 2019, Shigeru Hosono 2012). Unifying the software development process, the software installation and configuration should not be error-prone. The IT system configuration and management tool are used mostly to enable DevOps, it automates the process of system configuration. The configuration of the system is managed as rules which are organized as modules and classes (Diomidis Spinellis 2012, Rodríguez et al., 2013). The ITIL Framework is also an IT service management tool used in DevOps (Anthony Orr 2012). The architecture of ITIL was shown in figure 5.

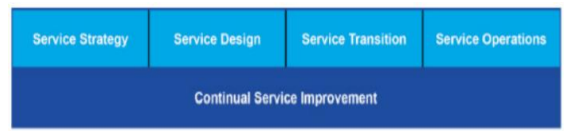

**Fig. 5.** ITIL architecture (Anthony Orr 2012)

**4. Challenges in Adopting DevOps:**

The adoption of DevOps in an IT firm is not so easy. In a company it's hard to change the culture of the organization and the practices involved in DevOps will not suit all circumstances (Leah et al., 2016, Khan et al., 2020). The challenges involved can be categorised as lack of awareness, support, technological feasibility and adapting to the change (Bucena and Marite 2017). Apart from organizational culture challenges, the other challenges include implementing in the existing process, architectural challenges, lack of automation for continuous testing and Legacy systems (Chen et al., 2017). The infrastructure also plays an important role in DevOps adoption and it should be compatible and lightweight (Khan et al., 2020). The challenges identified using the Fuzzy TOPSIS approach in (Rafi et al., 2020) were Data heterogeneity, data integration, error and inconsistent data, a misspelling in data entry, missing information, traceability of data, Data harmonization, visualisation of Data etc. The conflicts arise between the Development and Operations teams while deploying the newly developed features and fixing the problems. The development team develops a new feature without knowing the problems with the older version. The operations team has to fix the problem with the development team but the development team is ready to deploy a new One (Michael Hüttermann 2012). Some of the obstacles in software development for a highly regulated environment are 1) the operator doesn't know about the code 2) all artefacts are stored in a single isolated container 3) project must be approved before commencement 4) Poor collaboration (Morales et al., 2018).

## 5. Proposed Models:

A three-step model for DevOps adoption was proposed in (Luz et al., 2019). The model was evaluated by practical implementation. The relationship between the category's agility, automation, collaborative culture, continuous measurement, quality assurance, resilience, sharing and transparency was required for DevOps adoption (Luz et al., 2019). The iObserve approach used to tackle the challenges in DevOps adoption was discussed in (Heinrich et al., 2017). The iObserve approach follows MAPE (Monitor, Analyse, Plan, Execute) Control loop. The framework of iObserve approach is shown in figure 6.

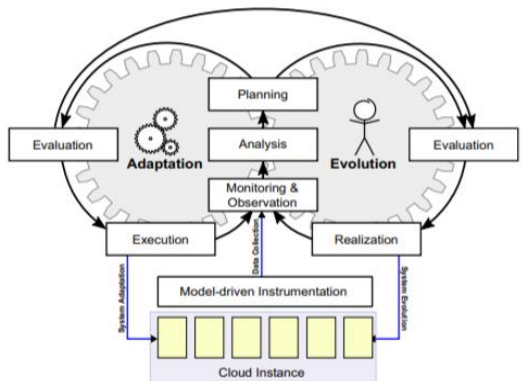

**Fig. 6.** iObserve model (Heinrich et al., 2017)

A technology transfer model shown in figure 7 was proposed in (Mikkonen et al., 2018) and it was proposed for a scalable multi-party organisation, academy and industry, multi-party team. It was reported in (Soha Solouki 2020) that DevOps performance increases with increase in integrity between DevOps and Knowledge management. The volunteer knowledge sharing should be encouraged in the companies.

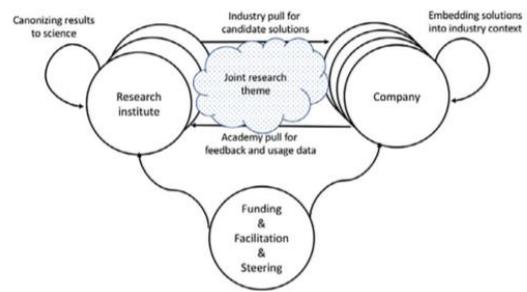

**Fig. 7.** Technology Transfer Model (Mikkonen et al., 2018)

A DevOps Maturity model was proposed in (Samer Mohamed 2015), with 5 levels of maturity and 4 dimensions of assessment. The proposed model is shown in figure 8 based on the CMMI model.

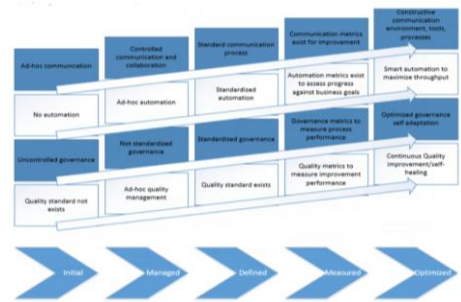

**Fig. 8.** DevOps Maturity Model (Samer Mohamed 2015)

The concept of Composable DevOps architecture was proposed in (McCarthy and Lorraine 2015). The composable nature is to break up things into smaller fragments and then to integrate the entire development pipeline. The composable DevOps is an iterative way to make collaboration between development and Operations.

## 6. Quality:

In software product development using new techniques, the quality of the developed product should also be maintained properly. The quality assurance of the product is improved by the use of an automated DevOps pipeline. The releases are made more often to improve the features of the product (Leah et al., 2016). The set of metrics used to assess and evaluate the DevOps practices were discussed in (Leah et al., 2016). The proposed organisational metrics to track the continuous improvement are Truck-Factor, Socio-Technical Congruence, Core-Periphery Ratio, Community Member Turnover and Smelly-Quitters. The Technical metrics are Lines of Code, Coupling Between Object Classes, – Code Change Process, developer-related Factors, Runtime Maintainability Measures and Operations Factors. In a fraud detection system designed in (Perez-Palacin et al., 2017), the Simtool was used to evaluate the metrics. The Fuzzy TOPSIS approach was used for the quality assessment in DevOps (Rafi et al., 2020).

## 7. Future works on DevOps:

By changing the configuration of the existing model, the upcoming model can be developed as multi-dimensional (Arulkumar and Lathamanju 2019). The DevOps team will possess desirable qualities or skills, full-stack development, analysis, functional, decision-making, social, testing, and advisory skills (Wiedemann et al., 2018). In the future, most of the IT firms will adopt the continuous deployment strategy (Rahman et al., 2015). The new metrics will be found to evaluate the quality of the DevOps development cycle (Leah et al., 2016). Tools like agile consolidator aimed at improving the quality of the process should be built (Doukoure et al., 2018). The auto-scaling policies in the combination of both container and node level will be developed (Kiyana Bahadori and Tullio Vardanega 2018). Some future works that can be done on working with Docker containers are Networking, Eclipse Plugin, Port management and Clustering capabilities (Robert Sandoval 2016). External clients are eager to collaborate with focused groups inside the solution providing enterprises to develop mixed solution assets in an agile fashion (McCarthy and Lorraine 2015).

## 8. Conclusion:

In this paper, we have discussed the building blocks of DevOps, challenges in adopting DevOps and the future work. DevOps is an emerging optimal practice that should be followed in a software development life cycle to increase releases, reliability, faster updates, effective use of customer feedback, increased quality assurance and security. It was clear that DevOps cannot be implemented in an existing pipeline on a single step. It should be implemented by small increments. The automation should be done in the repetitive tasks and at which humans tend to make errors. The cloud is the key player in DevOps, which takes part in integration, continuous delivery, security and in collecting feedback. Many frame-work of the architecture has been discussed and all these models rectified some challenges or improved the existing benefits of DevOps. Apart from all these factors, the team should be skilled and flexible enough to adapt the cultural changes.